# Einstein's Clocks and Langevin's Twins

Galina Weinstein[*]

In 1905 Einstein presented the Clock Paradox and in 1911 Paul Langevin expanded Einstein's result to human observers, the "Twin Paradox". I will explain the crucial difference between Einstein and Langevin. Einstein did not present the so-called "Twin Paradox". Later Einstein continued to speak about the clock paradox. Einstein might not have been interested in the question: what happens to the observers themselves. The reason for this could be the following; Einstein dealt with measurement procedures, clocks and measuring rods. Einstein's observers were measuring time with these clocks and measuring rods. Einstein might not have been interested in so-called biology of the observers, whether these observers were getting older, younger, or whether they have gone any other changes; these changes appeared to be out of the scope of his "Principle of relativity", or kinematics. The processes and changes occurring within observers seemed to be good for philosophical discussions. Later writers criticized Einstein's clock paradox. Einstein quickly replied with witty, smart and clever retorts.

**1. Time Dilation**

In the relativity paper Einstein derived the time dilation formula. He "imagines one of the clocks that can mark the time $t$ when at rest relative to the rest system ($K$), and the time $\tau$ when at rest relative to the moving system ($k$). This clock is located at the origin of the system $k$, and it is adjusted to mark the time $\tau$". Einstein asks: "what is the rate of this clock, when viewed from the rest system $K$?"

According to the Lorentz transformation[1] equation for the time,

$$\tau = \beta\left(t - \frac{vx}{c^2}\right),$$

where, $x = vt$ and,

$$\beta = \frac{1}{\sqrt{1 - v^2/c^2}}$$

Einstein obtains:

---

[*] Written while at the Center for Einstein Studies, Boston University

$$\tau = t\sqrt{1 - v^2/c^2} = t - \left(1 - \sqrt{1 - v^2/c^2}\right) t.$$

Einstein concludes that the time marked by the clock (viewed in the system at rest) is slow by $\left(1 - \sqrt{1 - v^2/c^2}\right)$ seconds, or to first order, by ½$v^2$/$c^2$ seconds.[2]

## 2. Clock Paradox

From this Einstein derived a "peculiar consequence":[3]

"If at the points *A* and *B* of *K* there are clocks at rest which, considered from the system at rest, are running synchronously, and if the clock at *A* is moved with the velocity *v* along the line connecting *B*, then upon arrival of this clock at *B* the two clocks no longer synchronize, but the clock that moved from *A* to *B* lags behind the other which has remained at *B* by ½t$v^2$/$c^2$ sec. (up to quantities of the fourth and higher order), where *t* is the time required by the clock to travel from *A* to *B*".

The above "peculiar consequence" came to be known as "the Clock Paradox".

John Stachel explained that, the most new important feature of time to emerge from the special theory of relativity is the clock paradox: a comparison of one clock in a moving system with many clocks in the rest system; there is no reciprocity, but one-many relationship. The clock paradox that was predicted by Einstein embodies the difference between Einstein and pre-relativistic electrodynamics of moving bodies.[4]

Einstein proposed an experimental test for the clock paradox: "From this we conclude that a balance-wheel clock that is located at the Earth's equator must run very slightly slower than an absolutely identical clock, subjected to otherwise identical conditions, that is located at one of the Earth's poles".[5]

## 3. Twin Paradox

Einstein spoke of clocks, while in 1911 Paul Langevin extended Einstein's Clock Paradox to human observers and the aging effect. Human observers, "twins", come instead of Einstein's clocks, and Langevin spoke of the "Twin Paradox". I will explain the crucial difference between Einstein and Langevin now.

Einstein *did not* present the so-called "Twin Paradox". Later Einstein continued to speak about the clock paradox. Einstein might not have been interested in the question: what happens to the observers *themselves*. The reason for this could be the following; he dealt with measurement procedures, clocks and measuring rods and he considered his theory as "a theory of measuring rods and clocks". In 1946 (published in 1949) Einstein wrote in his *Autobiographical Notes*: "The insight fundamental for the special theory of relativity is this: The assumptions (1) and (2) [the light postulate and the principle of relativity] are compatible if relations of a new type ('Lorentz transformation') are postulated for the conversion of coordinates and times of events.

With the given physical interpretation of coordinates and time, this is by no means merely a conventional step but implies certain hypotheses concerning the actual behavior of moving measuring rods and clocks, which can be experimentally confirmed or disproved".[6]

Einstein's observers were measuring time with clocks and measuring rods. Einstein might not have been interested in so-called biology of the observers, whether these observers were getting older, younger, or whether they have gone any other changes; these changes appeared to be out of the scope of his "Principle of relativity", or kinematics.

In Langevin's talk, "Evolution of Space and Time" ("L'Evolution de l'espace et du temps") given on April 11, 1911 Langevin imagines two arbitrary events in the history of an element of matter. Their time interval is measured by observers in non-uniform motion who have constantly moved together with the element. This time interval will be shorter than the one for a uniformly moving reference system. This latter reference system can be such that the two events considered could be situated at the same point, relative to which an element of matter has moved in a closed path, exactly as it is in Einstein's original clock paradox, and returns to its original position, due to its non-uniform movement. Langevin proves, "*for observers at rest relative to that element of matter, the time which has elapsed between the beginning and end of the path, the proper time* [in Minkowski's sense] *of the element of matter, will be shorter than for the observers in the uniformly moving reference system*".[7]

In other words, concludes Langevin, "the element of matter will have aged less between the beginning and end of its path than if it had not been accelerated, if it had instead remained at rest in a uniformly moving reference system".[8] And thus, "It follows from the result stated above, that the one having aged less is the one whose motion during the separation was furthest removed from uniform motion, the one most strongly accelerated".[9]

And immediately thereafter Langevin describes the famous twin paradox: "This remark gives a way, for any of us, who is willing to devote two years of his life, of knowing how the Earth will be in two hundred years time, to explore the future of the Earth, by hopping forward in the history of the latter, of two centuries, corresponding in his own life to only two years; but without any hope of return, without the possibility to come and inform us of the result of his journey, because any similar attempt can only throw him further and further into the future.

It is sufficient for this that our traveler agrees to shut himself up in a projectile, sent away from the Earth, with a speed sufficiently close to that of light, although less, which is physically possible, arranging that an encounter occurs with, for example, a star, after one year in the life of the traveler, that sends the spaceship back towards the Earth with the same speed. Returning to Earth, having aged by two years, he will climb out of his vehicle and find our globe aged by at least two hundred years, if his speed had stayed within an interval of less than twenty-thousandth of the speed of light. Solidly established experimental facts of physics allow us to state that the situation will really be as the one just described".[10]

**4. Clock Paradox and Superluminal Velocities**

Soon after Langevin's paper, Max Laue discussed the *clock* paradox thought experiment. In December 1911, Laue gave a modification to the paradox in his paper entitled, "Two Objections to the Theory of relativity and its Refutation". Laue meant two authors raise one objection to the theory: The objections were that, Einstein's clock paradox and Langevin's version of it as a twin paradox brought scholars to think that it might violate special relativity; because they thought that Einstein's peculiar clock consequence might enable superluminal signal velocities.[11]

Laue wrote the paper in order to answer the critics and explain to them why there can be no superluminal signal velocities. Yet Laue adhered to *Einstein's original version*, clocks and not human observers: A group of equivalent clocks, which were physically exactly the same, were moving in uniform velocity toward the same direction. The clocks moved so close one to another that they could be regarded as being simultaneously together at the same point. They all measured the proper time in Minkowski's sense on their corresponding world line. At a certain moment, they took separate paths, and after a certain time they came back again together to the same starting point in space. Afterwards they again went through a common movement as before. At the meeting point (between the places of separation and clock reunification), the clock that stayed in a "non-legitimate" frame of reference (not always inertial) lagged behind the other clock. This contradiction, says Laue "had prompted recently two authors at this point to raise one objection".[12]

During 1907 Einstein discussed with Wilhelm Wien the occurrences of velocities exceeding the speed of light in dispersive and absorptive media.[13] Emil Wiechert had been troubled with this issue for a long time and in his 1911 paper "Relativitätsprinzip und Äther" was still occupied by superluminal velocities.[14]

Wiechert thought that if superluminal velocities were possible within the scope of the "Relativity Principle" then all reference systems could not be equivalent. The systems of reference are equivalent insofar as we are dealing with a theory in which superluminal velocities are forbidden. If superluminal velocities are possible within the scope of the "Relativity Principle", then Wiechert sees in Einstein's clock paradox, or Langevin's version of Einstein's 1905 clock paradox, a contra example to the regular order of cause and effect; this is so, because if velocities exceed the speed of light, then some frames of reference see the returning twin younger, while the others may see him older. Wiechert thought there can be no agreement among the different clocks in the group in Laue's thought experiment. This was not permissible in a theory in which velocities do not exceed that of light. The relativity principle should absolutely deny the occurrence of superluminal velocities, and thus all systems should be completely equivalent.

Max Laue explained that Einstein and Langevin cancel the possibility of superluminal velocities, because cause comes before effect.[15] All agree that the returning twin is younger, and the twin who stayed at home is older. And that is the crux of Einstein's "eigentümliche Konsequenz".

Laue concluded that no new facts as to the existence of superluminal velocities were detected by experiment. Hence since there was no experimental evidence in favor of superluminal velocities, then at the time being there was no contradiction with the Principle of Relativity. This does not mean that the question is not interesting; on the contrary, it seems to Laue very interesting, but it should be dealt from the philosophical point of view. And hence it is surely to be treated with philosophical methods. Laue does not say that explicitly, but it seems that he thinks that the objections to the Einstein/Langevin thought experiment should be discussed by philosophical methods as well.[16]

**5. Anti-Semites against Clock Paradox**

The clock paradox was also an excuse for anti-Semites to blame the theory of relativity as an anti-German science and blame its author as well. In Berlin, Ernst Gehrcke, Philipp Lenard, and Paul Weyland advocated an anti-relativity propaganda campaign after 1916.[17] In November 1918 Einstein answered the first two by a Galilean dialogue between a relativist and a critic of the theory of relativity, "Dialogue about Objections to the Theory of Relativity" ("Dialog über Einwände gegen die relativitästheorie"); Einstein's imaginary character, "Kritikus", raises as claims against relativity the original clock paradox and not the twin paradox:[18]

**Kritikus**: "People like me have quite often expressed doubts of the most varied kind about the theory of relativity in journals; but rarely has one of you relativists responded. […] Since I see your willingness I come right to the matter at hand. Since the special theory of relativity has been formulated, its result of the delaying influence of motion upon the rate of clocks has elicited protest and, as it seems to me with good reasons. This result seems necessarily to lead to a contradiction with the very foundations of the theory".

Kritikus considers a Galilean coordinate system K in the sense of the special theory of relativity, that is, a frame of reference, relative to which isolated, material points move in straight lines and uniformly. Also, $U^1$ and $U^2$ are two identical clocks that are free from outside influences. These run at the same pace when they are in close proximity and also at any distance from each other, if they are both at rest relative to K. However, if one of the clocks, for example $U^2$, relative to K, is in a state of uniform translational motion, then according to the special theory of relativity it

should – as perceived from coordinate system K – go at a slower pace than the clock $U^1$ that is at rest relative to K. This result is in itself highly peculiar.

Kritikus then says that this result entails serious consequences if you look at the following well-known thought experiment. Let A and B be two distant points of the system K. A is the origin of K, and B is a point on the positive x-axis. The two clocks are initially at rest at point A. They run at the same pace, and the positions of the hands are the same. We now impart to clock $U^2$ a constant velocity in the positive direction of the x-axis, so that it moves towards B. At B we imagine the velocity reversed, so that clock $U^2$ returns to A. As it arrives at A, the clock is decelerated so that it is once again at rest relative to $U^1$; clock $U^2$ lags behind $U^1$: [19]

"According to the principle of relativity, the entire process must occur in exactly the same way if it is represented in a coordinate system K', that is co-moving with clock $U^2$. Then relative to K' it is clock $U^1$ that is moving, with clock $U^2$ remaining at rest. It then follows that $U^1$ should finally run behind $U^2$, in contradiction with the above result. Even the most devout adherents of the theory cannot claim that of two clocks, resting side by side, each is late relative to the other".

**Relativist**: Your last assertion is, of course, incontestable. But the entire line of reasoning is not legitimate because according to the special theory of relativity the coordinate systems K and K' are not equal. In fact, indeed, this theory claims only the equivalence of all Galilean (non-accelerated) systems, i.e., coordinate systems relative to which sufficiently isolated material points move uniformly in straight lines. The coordinate system K has certainly been such [a system], but not the temporarily accelerated system K'. One can therefore conclude that the clock $U^2$, after leaving away $U^1$ does not contradict the basis of the theory.

[…] **Kritikus**: […] In fact, it now seems not improbable to me that the theory may have no internal contradictions; but this is not sufficient to take it into serious consideration, *I just cannot see why one should be willing to take on such horrible complications and mathematical difficulties merely for the sake of an intellectual preference, namely for the idea of relativity*. In your last answer you yourself showed clearly enough that they are not minor.

**Relativist**: For several reasons we must willingly accept the complications to which the theory leads us. […] The following counterexample will show how inadvisable it is to appeal to so-called common sense as an arbiter in such things, Lenard himself says: so far no pertinent objections have been found to the validity of the *special* principle of relativity (i.e., the principle of relativity between uniformly translational motions of coordinate systems).[20] The uniformly moving train could as well be seen 'at rest' and the tracks, including the landscape, as 'uniformly moving'. Will the 'common sense' of the locomotive engineer allow this? He will object that he does not go on to heat and grease the *landscape* but rather the locomotive, and that consequently it must be the latter whose motion shows the effect of his labor".

*I wish to thank Prof. John Stachel from the Center for Einstein Studies in Boston University for sitting with me for many hours discussing special relativity and its history. Almost every day, John came with notes on my draft manuscript, directed me to books in his Einstein collection, and gave me copies of his papers on Einstein, which I read with great interest*---

[1] The system of values $x$, $y$, $z$, $t$ completely define the place and time of the event in the system $K$ at rest and the system of values $ξ, η, ζ, τ$ determine the event relative to the system k moving with respect to k with velocity v.

[2] Einstein, Albert, "Zur Elektrodynamik bewegter Körper, *Annalen der Physik* 17, 1, 1905, pp. 891-921, p. 904.

[3] Einstein, 1905, pp. 904-905.

[4] Stachel, John, "A world Without Time: the Forgotten Legacy of Gödel and Einstein", *Notices of the American Mathematical Society* 54, 2007, pp. 861-868, p. 866.

[5] Einstein, 1905, pp. 904-905.

[6] Einstein, Albert ,"Autobiographical notes" In Schilpp, Paul Arthur (ed.), *Albert Einstein: Philosopher-Scientist*, 1949, La Salle, IL: Open Court; pp. 1–95; pp. 52-53 and pp. 56-57.

[7] Langevin, Paul, "L'evolution de l'espace et du temps", *Scientia* 10, 1911, pp. 31-54, p. 48.

[8] Langevin, 1911, pp. 48-49.

[9] Langevin, 1911, p. 49.

[10] Langevin, 1911, p. 50.

[11] Superluminal phase velocities are allowed in electromagnetic medium.

[12] Laue, von Max, "Zwei Einwände gegen die Relativitätstheorie und ihre Widerlegung", *Physikalische Zeitschrift* 13, 1912, pp. 118-120; pp. 118-119.

[13] Einstein to Wien, 23 August or July, 1907, *The Collected Papers of Albert Einstein. Vol. 5: The Swiss Years: Correspondence, 1902–1914* (*CPAE*, Vol. 5), Klein, Martin J., Kox, A.J., and Schulmann, Robert (eds.), Princeton: Princeton University Press, 1993, Doc. 49; Einstein to Wien, 25 August or July, 1907, *CPAE*, Vol. 5, Doc. 50; Einstein to Wien, 29 July, 1907, *CPAE*, Vol. 5, Doc. 51; Wien to Einstein to Wien, 26 August, 1907, *CPAE*, Vol. 5, Doc. 55. See my paper discussing superluminal velocities in this ArXiv.

[14] Wiechert, Emil, Relativitätsprinzip und Äther, *Physikalische Zeitschrift* 12, 1911, pp. 689- 707.

[15] Laue, 1912, p. 119.

[16] Laue, 1912, p. 120; the twin paradox hosted dozens of papers and articles over the years, notable in his objections to the paradox was Herbert Dingle; many objections denied the asymmetry between the twins, and centered upon the erroneous central claim that according to the principle of relativity, one frame of reference sees the twin who took the journey to the star younger, while other frames may see him older.

Dingle's first paper against relativity was published during Einstein's lifetime: Dingle, Herbert, "The Relativity of Time", *Nature*, November, 1939, pp. 888-890.

Einstein owned two copies of Dingle's books in his personal library: Dingle, Herbert, *Science and human experience*, 1932, New York: Macmillan, and: Dingle, Herbert, *Through science to philosophy*, 1937, Oxford: Clarendon Press.

After Einstein's death Dingle became an ardent admirer of Newtonian mechanics and a bitter criticizer of Einstein's relativity. Dingle, Herbert, "Relativity and Space Travel", *Nature*, April 28, 1956, pp. 782-784; reply by McCrea, W. M., *Nature*, April 28, 1956, pp. 784-785; Dingle, Herbert, "Relativity and Space Travel", *Nature*, September 29, 1956, pp. 680-681; "Prof. Dingle's statements about the work of others are irresponsible" by McCrea, *Nature*, September 29, 1956, pp. 681-682; Dingle, Herbert, "The 'Clock Paradox' of Relativity", *Nature*, June 15, 1957, pp. 1242-1243.

A follower of Dingle: Sachs, Mendel, "A resolution of the Clock Paradox", *Physics Today*, September 1971, pp. 23-29. An answer to Sachs' paper by James Terrell: "The Clock 'paradox' – majority view", letters, *Physics Today*, January 1972, p. 9.

In 1971 four cesium beam atomic clocks were flown on regularly scheduled commercial jet flights around the world twice, once eastward and once westward, to test Einstein's clock paradox with macroscopic clocks. After a nonstop equatorial circumnavigation of the earth at various altitudes the predicted relativistic time was obtained for the flying clock. Hafele, J. C., and Keating, Richard. E., "Around-the-World Atomic Clocks: Predicted Relativistic Time Gains", *Science* 177, 1972, pp. 166-177.

Later that same year, the elderly Dingle published his strongest attack on Einstein's relativity: Dingle, Herbert, *Science at the Crossroads*, 1972, London: Martin Brian & O'Keeffe; and six years later in 1978 he passed away.

[17] Rowe, David E., and Schulmann Robert, *Einstein on Politics: His Private Thoughts and Public Stands on Nationalism, Zionism, War, Peace, and the Bomb*, 2007, Princeton: Princeton University Press, p. 11.

[18] Einstein, Albert, "Dialog über Einwände gegen die relativitästheorie", Die *Naturwissenschaften 6*, 1918, pp. 697-702, *The Collected Papers of Albert Einstein. Vol. 7: The Berlin Years: Writings, 1918–1921* (*CPAE*, Vol. 7), Janssen, Michel, Schulmann, Robert, Illy, Jószef, Lehner, Christoph, Buchwald, Diana Kormos (eds.), Princeton: Princeton University Press, 1998, Doc. 13; Rowe and Schulmann, 2007, p. 96.

[19] Einstein, 1918, *CPAE*, Vol. 7, Doc. 13; Rowe and Schulmann, 2007, p. 96.

[20] Einstein wrote the paper after he created the General Theory of Relativity. He therefore emphasized that he was discussing the special principle of relativity and not the general principle of relativity. This was important, because the twin paradox was also explained within the scope of General Relativity, but in this 1918 paper Einstein supplied only the special relativistic explanation.